\documentclass[onecolumn,draftcls]{IEEEtran}
\ifCLASSINFOpdf
\else
\fi

\usepackage{amssymb,amsfonts,amsmath}
\usepackage{algorithm,algorithmic}
\usepackage{graphicx}
\usepackage{epstopdf}
\usepackage{cite}
\begin{document}
%
\title{Sparsest Error Detection via Sparsity Invariant Transformation based $\ell_1$ Minimization}
%
%
%

\author{Suzhen Wang,
        Sheng Han,
        Zhiguo Zhang,
        and Wing Shing Wong,~\IEEEmembership{~Fellow,~IEEE}
\thanks{S. Wang and W. S. Wong are with the Department of Information Engineering, The Chinese University of Hong Kong, Hong Kong
(e-mail: ws012@ie.cuhk.edu.hk; wswong@ie.cuhk.edu.hk).}
\thanks{S. Han is with the Department of Electrical and Electronic Engineering, The University of Hong Kong, Hong Kong.}
\thanks{Z. Zhang was with the Department of Electrical and Electronic Engineering, The University of Hong Kong, Hong Kong. He is now with the School of Chemical and Biomedical Engineering, Nanyang Technological University, Singapore.}}
\maketitle

\begin{abstract}
This paper presents a new method, referred to here as the {\em sparsity invariant transformation based $\ell_1$ minimization}, to solve the $\ell_0$ minimization problem for an over-determined linear system corrupted by additive sparse errors with arbitrary intensity. Many previous works have shown that $\ell_1$ minimization can be applied to realize sparse error detection in many over-determined linear systems. However, performance of this approach is strongly dependent on the structure of the measurement matrix, which limits application possibility in practical problems. Here, we present a new approach based on transforming the $\ell_0$ minimization problem by a linear transformation that keeps sparsest solutions invariant. We call such a property {\em a sparsity invariant property} (SIP), and a linear transformation with SIP is referred to as {\em a sparsity invariant transformation} (SIT). We propose the SIT-based $\ell_1$ minimization method by using an SIT in conjunction with $\ell_1$ relaxation on the $\ell_0$ minimization problem. We prove that for any over-determined linear system, there always exists a specific class of SIT's that guarantees a solution to the SIT-based $\ell_1$ minimization is a sparsest-errors solution. Besides, a randomized algorithm based on Monte Carlo simulation is proposed to search for a feasible SIT.
\end{abstract}

\begin{IEEEkeywords}
Sparsest error detection, sparsest recovery, sparsity invariant transformation, SIT-$\ell_1$ minimization.
\end{IEEEkeywords}

%
\IEEEpeerreviewmaketitle

\section{Introduction}
This paper presents a new approach to transform the exact sparsest error detection problem for a linear system corrupted by additive errors with arbitrary intensity.
Let $y\in R^n$ denote a measurement vector, $A\in R^{n\times r}$ a measurement matrix, $x\in R^ r$ a hidden signal, and $e$ a noise vector. Then a noisy linear measurement system can be represented by the equation
\begin{equation}
\label{Basic Formulation}
y=Ax+e.
\end{equation}
Depending on the application, many such models have been proposed and analyzed \cite{burg2005vlsi, candes2005decoding,donoho2013high}. Here, we consider $e$ as an error vector and focus on an over-determined measurement system corrupted by sparse errors. In other words, we assume $n>r$, most entries of $e$ are zeros, and our goal is to detect the sparsest errors and thus determine a sparsest-errors solution.
The sparsest error detection problem can be formulated as the $\ell_0$ minimization problem as below:
\begin{equation}
\begin{split}
\nonumber
(\ell_0)\quad \quad \min_x \|y-Ax\|_0.\\
\end{split}
\end{equation}
In the rest of the paper, we also refer to this $\ell_0$ minimization problem as Problem ($\ell_0$).
Define $x^*$ as a solution to this problem and let $e^* = y-Ax^*$ denote the corresponding sparsest error vector.
For simplicity, and without loss of generality, we assume the rank of matrix $A$ is equal to $r$, the column number of $A$.

Usually, problem ($\ell_0$) is hard to solve due to its non-convexity and combinatorial complexity in high dimensions. A commonly accepted solution is to apply $\ell_1$ relaxation to obtain an approximate solution to Problem ($\ell_0$). The $\ell_1$-relaxed problem can be written as:
\begin{equation}
\label{Original L1 Minimization}
\begin{split}
\nonumber
(\ell_1)\quad \quad \min_{x} \|y-Ax\|_1.\\
\end{split}
\end{equation}
We refer to this $\ell_1$ minimization problem as Problem ($\ell_1$) or direct $\ell_1$ minimization problem in the rest of the paper. Define $\hat{x}$ as a solution to Problem ($\ell_1$) and let $\hat{e} = y-A\hat{x}$ denote the corresponding $\ell_1$-minimal error vector.

Over the years, researchers have observed empirically that $\ell_1$ minimization, arising for example from the basis pursuit problem \cite{chen1998atomic}, or from the application of the $\ell_1$-norm penalty in model selection problem \cite{tibshirani1996regression}, tends to produce solutions containing many zero entries. Subsequently, it was proved in \cite{donoho2001uncertainty}
that if a signal has a sparse decomposition under a specially-structured matrix, then the sparse decomposition can be achieved by $\ell_1$ minimization.
More recently, various sufficiency conditions \cite{donoho2003optimally,gribonval2003sparse,donoho2006most,candes2005decoding, candes2008restricted} on the measurement matrix have been established to guarantee exact sparsity recovery via $\ell_1$ minimization. Roughly speaking, these are conditions requiring a measurement matrix to be either incoherent or near orthonormal. From this, we see that the performance of $\ell_1$ minimization in finding the sparsest solution relies on the structure of the measurement matrix.

To remove this reliance, this paper provides a new methodology to ensure solution equivalence between $\ell_0$ and $\ell_1$ minimization after imposing a linear transformation on the measurement vector space. The linear transformation is required to preserve the sparsity property of an original solution, and thus, is referred to as a sparsity invariant transformation (SIT). A major theoretical contribution of this paper is to show that such an SIT exists. As a result, the proposed methodology allows the application of $\ell_1$ minimization to a general, \textit{"condition-free"}, measurement matrix for sparsest error detection or sparsest signal recovery.
%

For simplicity, we derive the theoretical results based on the sparsest error detection model for an over-determined system. Since the task of finding a sparsest solution to an under-determined system can be equivalently converted into a sparsest error detection problem in an over-determined system \cite{candes2005decoding}, our results also apply to an under-determined system for sparsest recovery.
Even though we have proven the existence of an SIT, there is no known algorithm for constructing an SIT
in general.
Instead, we propose a heuristic randomized algorithm based on Monte Carlo simulation for its construction.
Numerical results in section 6 demonstrate that the proposed methodology is effective in sparsest error detection, especially in cases where direct $\ell_1$ minimization fails to detect the sparsest errors.

The rest of the paper is organized as follows: In section 2 we formally introduce the sparsity invariant transformation $\Phi$.
In section 3 we prove the existence of $\Phi$ for a general error detection problem. Section 4 describes how the proposed methodology can be applied to the sparse representation problem. Section 5 presents a randomized algorithm to construct $\Phi$. Section 6 highlights some numerical results to demonstrate the advantages of the proposed methodology. There is evidence to show the new approach works for cases where direct $\ell_1$ minimization performs poorly.

\noindent\textit{Notation}

We use $span(A)$ to denote the subspace spanned by columns of $A$. Let $span([A,y])^{\perp}$ denote the orthogonal complementary subspace of $span([A,y])$. We use $I_{r}\in R^{r\times r}$ to denote an $r\times r$ identity matrix.
Let $[n]=\{1,2,....,n\}.$  Let $\mathcal{S}\subset [n]$ (resp. $\mathcal{S}\subset [m]$), denote by $A(\mathcal{S}, \cdot)$ (resp. $A(\cdot, \mathcal{S}$ ) ) a sub-matrix formed by rows (resp. columns) of matrix $A$ indexed by elements in $\mathcal{S}$. If $\mathcal{S} = \{i\}$, then $A(i, \cdot)$ (resp. $A(\cdot, i)$ ) denotes the $i$-th row (resp. $i$-th column) of matrix $A$.
If $y$ is a vector, let $y(\mathcal{S})$ denote a sub-vector formed by entries of $y$ indexed by elements $\mathcal{S}$. If $\mathcal{S} =\{i\}$, then $y(i)$ denotes the $i$-th entry of $y$. Let $supp(e)$ denote the support set of $e$.

\section{Sparsity Invariant Transformation}
In general, a solution to Problem $(\ell_1)$ may not be a solution to Problem ($\ell_0$). Instead of imposing conditions on the measurement matrix to guarantee an equivalence between the two solutions as many researchers have done in compressed sensing, we set out to create an equivalence by transforming the original linear system by means of an invertible linear transformation.  The invertible condition is required in order to avoid losing any useful information during the transformation.
Assume $\Phi$ is an invertible matrix, then the transformed $\ell_0$ minimization problem can be written as:
\begin{equation}
\begin{split}
\nonumber
(SIT_0)\quad \quad \min_{x}\|\Phi y -\Phi Ax\|_0.\\
\end{split}
\end{equation}
Define $x^*_{\Phi}$ as a solution to Problem ($SIT_0$) and let $e^*_{\Phi} = \Phi y - \Phi A x^*_{\Phi}$ denote the corresponding sparsest error vector.
To make the transformation meaningful, we need to find a specific $\Phi$ to make $e^*_{\Phi}$ and $e^*$ equivalent in a sense. Formally stated, we require the linear transformation satisfies the following \textbf{sparsity invariant property} (SIP).

\noindent \textit{Definition:}
$\Phi$ is an SIP transformation if it is invertible and
there exist non-zero scaling factors $\lambda$ and $\mu$ such that $\lambda e^*_{\Phi}$ is a sparsest error vector to Problem ($\ell_0$) and $\mu e^*$ is a sparsest error vector to Problem $(SIT_0)$.
(Noted that if Problem ($\ell_0$) has a unique solution then $\mu$ is equal to $\frac{1}{\lambda}$.)

In the rest of the paper, we refer to Problem ($SIT_0$) as the SIT-$\ell_0$ minimization problem and $\Phi$ is an SIT if it is an SIP transformation.

It follows from the SIP definition that ignoring differences in magnitude, an SIT preserves the sparsest error vector of a linear system but it may change the $\ell_1$-minimal error vector. If the sparsest error vector and the $\ell_1$-minimal vector do not coincide in the original system, i.e., $y=Ax+e$, then we will show that an SIT can be applied to transform the original system into an equivalent representation in which the above two error vectors coincide.

To continue the analysis, we need to find matrices with the sparsity invariant property.
To this end, define a family of matrices, $\mathcal{R}$, by
\begin{equation}
\nonumber
\label{Rotation}
\begin{split}
\mathcal{R} =\{&\Phi\in R^{n\times n} \ |  \ \Phi(span([A,y]) = span([A,y]), \Phi f = f,\\ &
\forall f\in span([A,y])^{\perp}\}.
\end{split}
\end{equation}

Since any element $\Phi$ in $\mathcal{R}$ has zero null-space, $\Phi$ is invertible.
The following lemma 1 shows that if a matrix belongs to $\mathcal{R}$, then its inverse matrix also belongs to it.

\noindent\textbf{Lemma 1} If $\Phi$ belongs to $\mathcal{R}$, then $\Phi^{-1}$, the inverse of $\Phi$, also belongs to $\mathcal{R}$.

The proof is simple since $\Phi \in \mathcal{R}$, then we have
\begin{equation}
\nonumber
\Phi^{-1} f  = \Phi^{-1} \Phi f = f.
\end{equation}
for any $f\in span([A,y])^{\perp}$. Thus lemma 1 holds. $\blacksquare$

The next proposition proves that $\mathcal{R}$ defines a family of matrices with SIP.

\noindent\textbf{Proposition 1} If $\Phi\in \mathcal{R}$ then $\Phi$ satisfies the sparsity invariant property.

\noindent\textbf{Proof} Using notation introduced previously, let $e^*$ be a sparsest error vector to Problem $(\ell_0)$ and $e^*_{\Phi}$ be a sparsest error vector to Problem ($SIT_0$). Let $e_1 = \Phi^{-1}e^*$.
Since $e^*\in span([A, y])$ and $\Phi^{-1}\in \mathcal{R}$ according to lemma 1, it follows that $e_1\in span([A, y])$. Therefore, there exists a non-zero real number $\lambda$ such that
\begin{equation}
\nonumber
y-\lambda e_1 \in span(A).
\end{equation}
Denote $y-\lambda e_1 =Ax_1$, we have
\begin{equation}
\label{T2}
e^* = \Phi e_1 = \frac{1}{\lambda}\Phi(y-Ax_1),
\end{equation}
which implies
\begin{equation}
\label{Direction1}
\|e^*\|_0 \ge \|e^*_{\Phi}\|_0.
\end{equation}

On the other hand, let $ e_2 = \Phi e^*_{\Phi} \in span([\Phi A,\Phi y])$, then there exist a non-zero real number $\mu$ and a vector $x_2$ such that
\begin{equation}
\nonumber
\Phi y - \mu e_2 = \Phi A x_2.
\end{equation}
From this, it follows that
\begin{equation}
\label{T1}
e^*_{\Phi} = \Phi^{-1} e_2 = \frac{1}{\mu} (y-Ax_2),
\end{equation}
which implies
\begin{equation}
\label{Direction2}
\|e^*_{\Phi}\|_0 \ge \|e^*\|_0.
\end{equation}
According to $(\ref{Direction1})$ and $(\ref{Direction2})$, we infer that
\begin{equation}
\label{SparsityEquivalence}
\|e^*_{\Phi}\|_0 = \|e^*\|_0.
\end{equation}
Based on $(\ref{T2})$ and ($\ref{SparsityEquivalence}$), we know that $\lambda e^*$ is a sparsest error vector to Problem $(SIT_0)$. Based on $(\ref{T1})$ and ($\ref{SparsityEquivalence}$), we know that $\mu e^*_{\Phi} $ is the sparsest error vector to Problem ($\ell_0$). Thus proposition 1 holds. $\blacksquare$


\section{Exact Sparsest Recovery}

In the previous section, we present the sparsity invariant property (SIP) for a linear transformation $\Phi$, which preserves the sparsest error vector after transformation.
In this section, we aim to prove that for any $\ell_0$ minimization problem, an SIT always exists.  This ensures that a sparsest error detection problem can be equivalently converted into an SIT-based $\ell_1$ minimization problem defined by Problem ($SIT_1$):
\begin{equation}
\nonumber
\begin{split}
(SIT_1) \quad \quad \min_{x} \| \Phi y- \Phi Ax\|_1.\\
\end{split}
\end{equation}
Define $\hat{x}_{\Phi}$ as a solution to Problem ($SIT_1$) and let $\hat{e}_{\Phi}$ be the corresponding $\ell_1$-minimal error vector. We also refer to this problem as an SIT-$\ell_1$ minimization problem in the rest of the paper.

Before we prove the existence result, we introduce an example to demonstrate the essential concept of SIT-$\ell_1$ minimization in sparsest error detection.
\subsection{Example where direct $\ell_1$ minimization fails}
Researchers often use $\ell_1$ norm to relax $\ell_0$ norm in order to avoid NP-hard computational complexity. Unfortunately, in many cases direct $\ell_1$ relaxation fails to detect the sparsest errors. The following example presents one of the failed cases.
Define $A$ and $y$ by
\begin{equation}
\nonumber
A= [-1, \ 1, \ -10]^T \
\ {\rm and} \ \
y = [-1,\ 1, \ 0]^T.
\end{equation}
Then the corresponding sparsest error vector is
$e^* = [0,\ 0,\ 10]^T
$
and the $\ell_1$-minimal error vector is
$\hat{e}= [-1,\ 1,\ 0]^T.
$
Obviously, the two error vectors are not the same and they have different support sets.

To take the SIT-$\ell_1$ minimization approach we define a transformation $\Phi$ by
\begin{equation}
\nonumber
\Phi =
\begin{bmatrix}
0.5000 &   0.5000  &  0.7071\\
    0.5000 &   0.5000 &  -0.7071\\
   -0.7071   & 0.7071      &   0\\
\end{bmatrix}.
\end{equation}
One can verify that $span([A,y])^{\perp} = \{\lambda z|\lambda \in R\}$ where vector $z$ is set as
$[1,\ 1,\ 0]^T$,
and $\Phi z = z$ for the defined $\Phi$. This shows that $\Phi$ belongs to $\mathcal{R}$ and
hence it is an SIT.
Moreover,
\begin{equation}
\nonumber
\Phi y = [ 0,\ 0,\ 1.4142]^T
\ {\rm and} \
\Phi A= [7.0711,\ -7.0711, \ 1.4142]^T.
\end{equation}
And the sparsest error vector after transformation is
$e^*_{\Phi} = [0, \ 0,\ 1.4142]^T
$. Obviously $e^*_{\Phi} ( = 0.14142 e^*)$ is equal to the original sparsest error vector except the magnitude.
The $\ell_1$-minimal error vector after transformation is
$\hat{e}_{\Phi} = [0, \ 0,\ 1.4142]^T
$ which is equal to $e^*_{\Phi}$.  This example presents a case where SIT-$\ell_1$ minimization outperforms direct $\ell_1$ minimization in sparsity seeking.

Ignoring differences in magnitude, an SIT can be geometrically understood as a rotation on the measurement vector space 
around the "axis" defined by $span([A,y])^{\perp}$ to make the $\ell_1$-minimal and $\ell_0$-minimal errors coincide. The top figure in Fig.1 shows a failed case of directly applying $\ell_1$ minimization for sparsest detection whereas the bottom figure in Fig.1 presents the scenario that the $\ell_1$-minimal error vector coincides with the $\ell_0$-minimal error vector after applying an SIT.

\begin{figure}[t]
\begin{center}
   \includegraphics[width=10cm,height=15cm]{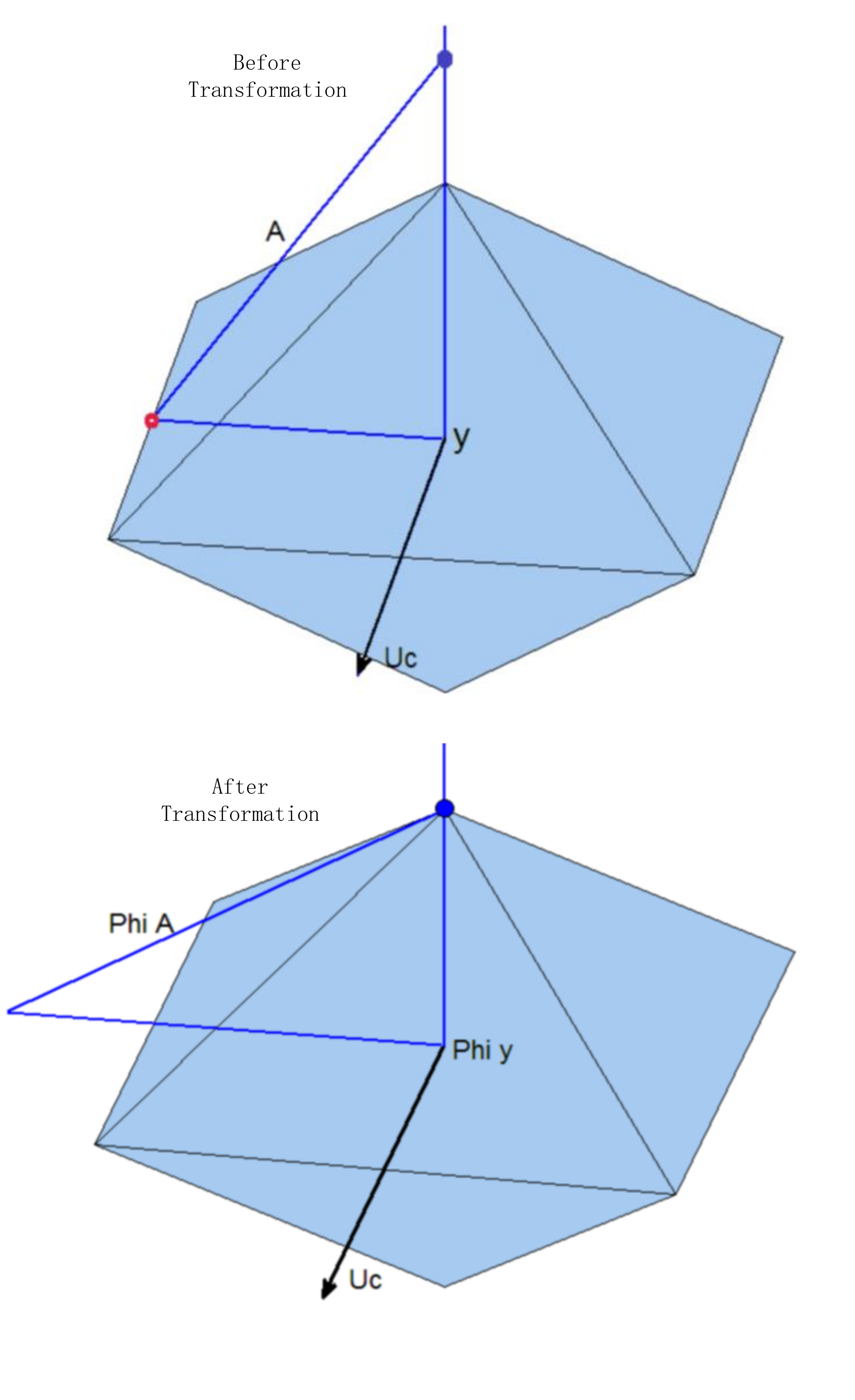}
\end{center}
\caption{$A$ denotes a measurement matrix, here it is a column vector, $y$ is the measured data, Phi A and Phi y denote $\Phi A$ and $\Phi y$ respectively, "axis" $U_c$ denotes $span([A,y])^{\perp}$, and the blue ball denotes an $\ell_1$ ball.
The red point denotes the $\ell_1$-minimal error vector and the blue point denotes the $\ell_0$-minimal error vector. The red point and the blue point are two different points before transformation, but they coincide after transformation. }
\end{figure}

\subsection{Mathematical Verification}
In this subsection, we present a proof of the existence of an SIT that can guarantee the detection of the sparsest errors for an
error detection problem. To facilitate subsequent discussions, we first introduce a family of polytopes, $\Gamma(t)$, defined as below:
\begin{equation}
\nonumber
\label{Slice}
\Gamma(t)= \{\alpha \ | \ \alpha = [A, y]\begin{bmatrix}
\beta\\
f
\end{bmatrix}, \ \beta \in R^{r}, \ f\in R > 0 ,\ \|\alpha\|_1\le t\}.
\end{equation}
From the definition of $\Gamma(t)$, one can infer the polytopes defined by $\Gamma(t)$ have the same shape and their scales vary with parameter $t$.

Assume $\alpha^*$ is one of the extreme points of $\Gamma(t)$, according to the separation theorem there exists a specific vector $\textbf{a}\in span([A,y])$ that guarantees:
\begin{equation}
\label{Support Condition}
\textbf{a}^T(\alpha-\alpha^*)<0, ~\forall \alpha \neq \alpha^*, \alpha\in \Gamma(t).
\end{equation}

To better understand SIT, we first show in lemma 2 that there are finitely many sparsest error vectors to Problem $(\ell_0)$.

\noindent \textbf{Lemma 2} The number of the sparsest error vector $e^*$ to Problem ($\ell_0$) is no more than $\dbinom{n}{r}$.

\noindent\textbf{Proof} Since we can always find a row index set $\mathcal{S}\in [n]$ with $rank( A(\mathcal{S}, \cdot) )= r$ such that
\begin{equation}
\nonumber
(y+e)(\mathcal{S}) = A(\mathcal{S}, \cdot)\bar{x}
\end{equation} no matter what the real error $e$ is.
Given such an $\mathcal{S}$, both $\bar{x}$ and the corresponding error vector are uniquely determined. There are at most $\dbinom{n}{r}$ such $\mathcal{S}$. Therefore, the number of the sparsest error vectors is no more than $\dbinom{n}{r}$, thus lemma 2 holds. $\blacksquare$

Based on lemma 2, we arrive at the following conclusion in lemma 3.

\noindent \textbf{Lemma 3} If the sparsest error vector $e^*$ to Problem ($\ell_0$) is not unique, then different sparsest error vector has different support set.

\noindent \textbf{Proof} Suppose $e_1$ and $e_2$ are two different sparsest error vectors that share the same support set. Assume
\begin{equation}
\begin{split}
\nonumber
&e_1 = y -Ax_1\\
&e_2 = y - Ax_2.
\end{split}
\end{equation}
Then we have
\begin{equation}
\nonumber
t e_1 +(1-t)e_2 = y -A[tx_1 +(1-t)x_2]
\end{equation}
for all real number $t$. Since $t$ can be an arbitrary real number, therefore we have infinitely many sparsest error vectors. This conclusion contradicts lemma 2, therefore the supposition is wrong, thus lemma 3 holds. $\blacksquare$

The following lemma characterizes another important feature of the sparsest error vectors to Problem ($\ell_0$).

\noindent \textbf{Lemma 4} If $e^*$ is one of the sparsest error vectors to Problem ($\ell_0$), then it is an extreme point of $\Gamma(\|e^*\|_1)$.

\noindent \textbf{Proof} If $e^*$ is not an extreme point, there exist two points $p_1$ and $p_2$ such that
\begin{equation}
\nonumber
e^* = tp_1 +(1-t)p_2,
\end{equation}
for some $t\in(0,1)$.
Therefore, we have
\begin{equation}
\label{Relationship}
\|e^*\|_1 = \|tp_1 +(1-t)p_2\|_1 \le t\|p_1\|_1 +(1-t)\|p_2\|_1.
\end{equation}
Since $e^*$ lies on the boundary of $\Gamma(\|e^*\|_1)$, we can replace the inequality in ($\ref{Relationship}$) with an equality. On the other hand, the equality in $(\ref{Relationship})$ holds if and only if the following two conditions are satisfied:
\begin{equation}
\nonumber
\|e^*\|_1 = \|p_1\|_1 = \|p_2\|_1,
\end{equation}
and the non-zero entries of $p_{1}$ and $p_{2}$ share the same signs.  The latter condition
implies $ supp(e^*) = supp(p_1)\cup supp(p_2)$. Since $e^*$ is a sparsest point of $\Gamma(t)$, we must have $supp(e^*) = supp(p_1) =supp(p_2)$. From the definition of $\Gamma(\|e^*\|_1)$, one can infer that with proper scaling parameters, denoted by $\lambda_1$ and $\lambda_2$, $\lambda_1 p_1$ and $\lambda_2 p_2$ are actually two sparsest error vectors to Problem ($\ell_0$). According to lemma 3, different sparsest error vector has different support set. Therefore, $\lambda_1 p_1$, $\lambda_2 p_2$ and $e^*$ must be the same sparsest error vector, that is,
\begin{equation}
\nonumber
e^* = p_1 = p_2.
\end{equation}
This result implies $e^*$ is an extreme point of $\Gamma(\|e^*\|_1)$. Thus lemma 4 holds. $\blacksquare$

According to lemma 4, we observe that $\frac{\|\hat{e}\|_1}{\|e^*\|_1}e^* $ is one of the extreme points of polytope $\Gamma(\|\hat{e}\|_1)$, where $\hat{e}$ is a sparsest error vector to Problem ($\ell_1$). Obviously, there exists a vector $\textbf{a}$ with the property:
\begin{equation}
\label{ReferHyperplane}
\textbf{a}^T(\alpha-\frac{\|\hat{e}\|_1}{\|e^*\|_1}e^*)<0, \ \forall \alpha\neq \frac{\|\hat{e}\|_1}{\|e^*\|_1}e^*,  \alpha\in \Gamma(\|\hat{e}\|_1).
\end{equation}
Condition $(\ref{ReferHyperplane})$ motivated us to  establish the following proposition below, which is essential to the proposed methodology.

\noindent \textbf{Proposition 2} If $\Phi \in \mathcal{R}$ is orthogonal and satisfies the equation
\begin{equation}
\label{SetPsi}
\Phi u_{r+1} = \textbf{a},
\end{equation}
with $\textbf{a}$ defined in ($\ref{ReferHyperplane}$) and $u_{r+1}$ being a normalized $l_2$ vector in $ span([A,y])$, orthogonal to $span(A)$, then
\begin{equation}
\nonumber
\begin{split}
e^*_{\Phi} = \hat{e}_{\Phi},
\end{split}
\end{equation}
where $e^*_{\Phi}$ and $\hat{e}_{\Phi}$ are an sparsest error vector and an $\ell_1$-minimal error vector after transformation respectively.

It should be noted that since both $\textbf{a} $ and $u_{r+1}$ belong to $span([A, y])$, there exist transformations to establish ($\ref{SetPsi}$).

\noindent \textbf{Proof}:
According to proposition 1, we have
\begin{equation}
\label{ReferProposition2}
e^* = \frac{\|e^*\|_1}{\|e^*_{\Phi}\|_1}e^*_{\Phi},
\end{equation}
for any $\Phi \in \mathcal{R}$.
By substituting $(\ref{ReferProposition2})$ into ($\ref{ReferHyperplane}$), we get
\begin{equation}
\label{SupportingPlane}
\textbf{a}^T(\alpha-\frac{\|\hat{e}\|_1}{\|e^*_{\Phi}\|_1}e^*_{\Phi})<0, ~\forall \alpha \neq \frac{\|\hat{e}\|_1}{\|e^*_{\Phi}\|}e^*_{\Phi}, \alpha \in \Gamma(\|\hat{e}\|_1).
\end{equation}
Setting $\theta = \frac{\|e^*_{\Phi}\|_1}{\|\hat{e}\|_1}\alpha$, we get
\begin{equation}
\nonumber
\textbf{a}^T(\theta - e^*_{\Phi})< 0, \ \forall \theta \neq e^*_{\Phi}, \theta \in \Gamma(\|e^*_{\Phi}\|_1).
\end{equation}
Observe that $\hat{e}_{\Phi}\in \Gamma(\|e^*_{\Phi}\|_1)$.  Suppose $ \hat{e}_{\Phi} \neq e^*_{\Phi}$
we have
\begin{equation}
\label{inequality}
\textbf{a}^T( \hat{e}_{\Phi}- e^*_{\Phi})< 0.
\end{equation}
Combining ($\ref{SetPsi}$) with ($\ref{inequality}$) we obtain the inequality
\begin{equation}
\begin{split}
\label{Contra1}
(\Phi u_{r+1})^T(\hat{e}_{\Phi} - e^*_{\Phi})<0.
\end{split}
\end{equation}
On the other hand, there exist vectors $Ax_1$ and $Ax_2$ such that
\begin{equation}
\begin{split}
\nonumber
\hat{e}_{\Phi} = \Phi y - \Phi Ax_1,\\
e^*_{\Phi} = \Phi y- \Phi Ax_2.
\end{split}
\end{equation}
It follows from
\begin{equation}
\nonumber
u_{r+1}^T A(x_1 -x_2)=0,
\end{equation}
that
\begin{equation}
\nonumber
\label{Contra2}
(\Phi u_{r+1})^T(\hat{e}_{\Phi} - e^*_{\Phi}) = 0,
\end{equation}
which contradicts the result in ($\ref{Contra1}$).
So we have
\begin{equation}
\nonumber
\begin{split}
e^*_{\Phi} = \hat{e}_{\Phi}.
\end{split}
\end{equation}
Thus proposition 2 holds. $\blacksquare$

Proposition 2 shows that for any $\ell_0$ minimization problem, there exists an SIT which enables the detection of the sparsest errors
by means of SIT-$\ell_1$ minimization of a corresponding problem.
In the following discussion, we reveal another nice property of SIT-$\ell_1$ minimization.

\noindent \textbf{Corollary 1} If $\Phi$ is an orthogonal matrix in $\mathcal{R}$ that satisfies ($\ref{SetPsi}$), then the corresponding SIT-$\ell_1$
minimization problem has a unique solution.

\noindent \textbf{Proof}
Suppose the corresponding SIT-$\ell_1$ problem does not have a unique solution. Given
there are two distinct solutions to the SIT-$\ell_1$ problem there will be two distinct $\ell_1$-minimal error vectors, $e_1$ and $e_2$, since $A$ is assumed to have full column rank.
For all $t$ in (0,1), define an error vector $e_t$ by
\begin{equation}
\nonumber
e_t = t e_1 + (1-t) e_2.
\end{equation}
It follows that
\begin{equation}
\nonumber
\|e_t\|_1\le t\|e_1\|_1 + (1-t) \|e_2\|_1.
\end{equation}
Since both $e_1$ and $e_2$ are $l_1$-minimal error vectors, we have
\begin{equation}
\nonumber
\|e_t\|_1 = \| e_1\|_1 =\|e_2\|_1,
\end{equation}
which implies that $e_t$ is also an $\ell_1$-minimal error vector to the SIT-$\ell_1$ minimization problem.
According to proposition 1 and 2, we know that for any $t \in (0,t)$, there exists a scaling factor, $\lambda$, that can scale
$\lambda e_t$ to become a sparsest error vector to Problem ($\ell_0$). However, this conclusion contradicts lemma 2, which states that the number of sparsest error vectors to Problem ($\ell_0$) is finite. By contradiction, it follows that corollary 1 holds. $\blacksquare$

This corollary implies that the uniqueness of the $\ell_1$-minimal error vector is guaranteed in the proposed SIT-based methodology.

\section{Sparsest Solutions for an Under-determined System by SIT-$\ell_1$ Minimization }
In this section, we consider finding sparsest solutions to an under-determined system,
\begin{equation}
\label{UnderdeterminedSystem}
\tilde{y} = Fe,
\end{equation}
where $\tilde{y}$ denotes measured data, $F$ denotes an $m\times n$ measurement matrix with $m<n$, and $e$ denotes an unknown signal to be estimated.
Although an under-determined system has many solutions, usually people are more interested in the sparsest one for its succinctness in capturing the essential features of a phenomenon. The sparsest solution can be obtained by solving the following combinatorial optimization problem:
\begin{equation}
\label{Sparsity Seeking}
\min_{e} \|e\|_0, ~s.t.~ \tilde{y} = Fe.
\end{equation}
To reduce the computational complexity, $\ell_1$ relaxation method was introduced to provide an approximate solution to problem ($\ref{Sparsity Seeking}$). The relaxed problem can be written as
\begin{equation}
\label{Basis Pursuit}
\min_{e} \|e\|_1, ~s.t.~ \tilde{y} = Fe.
\end{equation}

Recently, sufficient conditions have been proposed to guarantee the sparsest recovery ability of the above $\ell_1$ minimization problem.
However, as mentioned before, these conditions may limit the applicability of the $\ell_1$ relaxation technique to various practical problems.

To avoid imposing conditions on the measurement matrix $F$, we want to apply an SIT to achieve a solution equivalence between $\ell_0$ and $\ell_1$ minimizations in the under-determined setting.
To utilize previously derived results, we convert problem ($\ref{Sparsity Seeking}$) into an error detection problem. To start the conversion, we first compute a least-mean-squares solution, $ y = F^{+} \tilde{y}$, to the under-determined system $\tilde{y}=Fe$. $F^{+}$ denotes the pseudo-inverse of $F$. Assume the dimension of the kernel of $F$ is $r$. Then we introduce a matrix $A$ of $n\times r$, whose range is the kernel of $F$. Therefore, any vector belonging to the kernel of $F$ can be represented as $Ax$ with $x\in R^r$ a coefficient vector. Finally, as one can verify problem $(\ref{Sparsity Seeking})$ can be equivalently converted into an error detection problem as below:
\begin{equation}
\label{Converted Error Detection}
\min_x \| y - Ax\|_0.
\end{equation}
Observe that this problem is the same as Problem ($\ell_0$), therefore the corresponding SIT-$\ell_1$ minimization can be written as
\begin{equation}
\begin{split}
\nonumber
\min_{x} \|\Phi y -\Phi Ax\|_1,
\end{split}
\end{equation}
where $\Phi$ is an orthogonal SIT.
It follows from the SIT definition that $\Phi$ should satisfy
\begin{equation}
\nonumber
(con \ 1)\quad \quad \Phi f = f, \forall f \in span([A,y])^{\perp}.
\end{equation}
According to proposition 2, SIT-$\ell_1$ minimization problem succeeds to recover a sparsest solution if $\Phi$ further satisfy condition $(\ref{SetPsi})$, that is,
\begin{equation}
\nonumber
(con \ 2)\ \quad\quad \Phi u_{r+1} = \textbf{a}.\quad \quad \quad \quad \quad \quad \ \ \
\end{equation}
Since $y\in span(F^T)$ and $span(A) = span(F^T)^{\perp}$, the columns of $A$ are orthogonal to $y$.
As a result, $u_{r+1}$ in $(con\ 2)$ can be set as $u_{r+1}=\frac{y}{\|y\|_2}$.

On the other hand, as one can verify, the above SIT-$\ell_1$ minimization problem is equivalent to the following problem:
\begin{equation}
\label{SIT-underdetermined}
\min_{e} \|e\|_1, \ s.t. \ \tilde{y} =F\Phi^T e.
\end{equation}
Problem $(\ref{SIT-underdetermined})$ actually formulates an SIT based $\ell_1$ minimization for a problem of seeking a sparsest solution to an under-determined system.

In conclusion, a sparsest solution to problem ($\ref{Sparsity Seeking}$) can be determined by problem ($\ref{SIT-underdetermined}$) provided that an orthogonal $\Phi$ satisfies $(con \ 1)$ and $(con\ 2)$.
In fact, we can derive a vector decomposition representation of $\Phi F^T$ according to $(con \ 1)$ and $(con\ 2)$.
First expand matrix $F$ as follows:
\begin{equation}
\nonumber
F = [f_1,...,f_m]^T,
\end{equation}
where $f_i\in R^n$ denotes the transpose of the $i$-th row vector of $F$. Each $f_i$ is actually orthogonal to $span(A)$.
Since $Fu_{r+1}\neq 0$, we decompose each $f_i$ as follows:
\begin{equation}
\nonumber
\begin{small}
f_i = (f_i^T u_{r+1} ) u_{r+1} + [f_i - (f_i^T u_{r+1} )u_{r+1}], \ \forall i \in [m].
\end{small}
\end{equation}
Obviously, the second term belongs to $span([A,y])^{\perp}$, according to $(con\ 1)$, one gets
\begin{equation}
\nonumber
\Phi [f_i - (f_i^T u_{r+1} )u_{r+1} ] = [f_i - (f_i^T u_{r+1} )u_{r+1} ].\quad
\end{equation}
According to ($con\ 2$), one gets
\begin{equation}
\nonumber
\Phi f_i = (f_i^Tu_{r+1}) \textbf{a} + [f_i - (f_i^T u_{r+1})u_{r+1}]. \quad \quad \quad
\end{equation}
In short, we have
\begin{equation}
\label{Transform2}
\Phi F^T = diag(Fu_{r+1})\textbf{A} + F^T -diag(F u_{r+1})U,
\end{equation}
where $\textbf{A} =[\textbf{a},...,\textbf{a}]$ and $U = [u_{r+1},...,u_{r+1}]$, both $\textbf{A}$ and $U$ have $m$ identical columns.

From $(\ref{Transform2})$, we can see to the determination of $\Phi F^T$ can be reduced to a problem of determination of $\textbf{a}$. If the exact value of $\textbf{a}$ is derived in some way, then a sparsest solution can be easily obtained by solving the corresponding SIT-$\ell_1$ minimization problem.

\section{Randomized Algorithm Based on Monte Carlo}
Previous discussions imply that finding a sparsest solution via SIT-$\ell_1$ minimization is equivalent to determining a feasible choice of $\textbf{a}$. Unfortunately, we have not found a tractable way to obtain a feasible choice for $\textbf{a}$ so far. Instead we propose a randomized algorithm based on the Monte Carlo method to search for one. The randomized algorithm may be inefficient when the data dimension is high, but it allows for parallel computation for speeding up.

Let $U_r$ denote an orthobasis of $span(A)$. By applying a singular value decomposition on $A$ we get $A =U\Sigma V$. Since the rank of $A$ is assumed to be $r$, $U_r$ can be achieved by assigning the first $r$ columns of $U$ to it.
Assume $y\neq span(A)$, then there exists a unit vector $u_{r+1} $ that makes $U_{r+1} = [U_r, u_{r+1}] $ be an orthobasis of $span([A,y])$.

Noted that $\textbf{a}$ belongs to $span(U_{r+1})$ and has its $\ell_2$ norm equal to 1.
Denoted by $\mathcal{A}$ the set of feasible choices of $\textbf{a}$ that satisfy condition $(\ref{Support Condition})$. Let
\begin{equation}
\nonumber
\mathcal{B} = \{f\ | \ f=U_{r+1}^T \textbf{a},\|\textbf{a}\|_2 =1, \textbf{a}\in \mathcal{A}\}.
\end{equation}
If we generate an $f\in R^{r+1}$ with i.i.d. Gaussian entries and normalize its $\ell_2$ norm to one, then a random $\textbf{a}$ can be generated by the equation $\textbf{a} = U_{r+1}f$. Obviously, a random  $f$ belongs to $\mathcal{B}$ with a certain fixed probability.
If we generate enough $\ell_2$ normalized random $f$, the event that at least one of these random vectors will generate a feasible value of $\textbf{a}$ can happen with a high probability.

To ease the implementation, we recommend to convert the sparsest error detection problem into a problem of finding sparsest solutions to an under-determined system. Hence we can avoid dealing with $\Phi A$ since this term varnishes after conversion. The conversion procedure is as follows:
Since $U_{r+1}$ contains an orthobasis of $span([A,y])$, $y$ can be represented as:
\begin{equation}
\nonumber
y =  [U_r, u_{r+1}]\begin{bmatrix}
-\beta            \\
t
\end{bmatrix},
\end{equation}
with $t \in R$ and $\beta\in R^{r}$.
Moreover, for any $x\in R^r$, we have $Ax = U_r h$ for some $h\in R^r$.
As a result, Problem ($\ell_0$) can be re-written as:
\begin{equation}
\nonumber
\min_h \|[U_r, u_{r+1}]\begin{bmatrix}
-\beta            \\
t
\end{bmatrix} - U_r h\|_0.
\end{equation}
Set $z =h+\beta $, then the above problem can be further simplified to
\begin{equation}
\label{Orthogonified Problem}
\min_z \|tu_{r+1} - U_r z\|_0.
\end{equation}
On the other hand, set $F\in R^{(n-r)\times n}$ as
\begin{equation}
\label{F Form }
\nonumber
F = [u_{r+1},...,u_n]^T,
\end{equation}
where $\{u_{r+1},...,u_n\}$ denotes an orthobasis of $span(A)^{\perp}$.
So we have $FU_r = 0$ and thus we get
\begin{equation}
\nonumber
tFu_{r+1}-FU_{r}z = tFu_{r+1} = [t,0,...,0]^T.
\end{equation}
Then an equivalent representation of problem ($\ref{Orthogonified Problem}$) can be written as
\begin{equation}
\begin{split}
\label{DualForm2}
\min_{e} \|e\|_0, \ s.t. \ Fe =  [t,0,...,0]^T.
\end{split}
\end{equation}
Thus we finish the conversion.

According to the discussions in the last section, we can see, the corresponding SIT-$\ell_1$ minimization formulation to problem ($\ref{DualForm2}$) can be written as:
\begin{equation}
\nonumber
\min_{e} \|e\|_1, \ s.t.\ F\Phi^T e = [t,0,...,0]^T,
\end{equation}
where $\Phi$ is required to satisfy
\begin{equation}
\label{Set1}
\ \Phi u_{i} = u_{i}, i\in \{r+2,...,n\},
\end{equation}
and
\begin{equation}
\label{Set2}
\Phi u_{r+1} = \textbf{a}. \quad\quad\quad\quad\quad\quad
\end{equation}
Based on $(\ref{Set1})$ and $(\ref{Set2})$, we get
\begin{equation}
\nonumber
\Phi F^T = \Phi [u_{r+1},...,u_n] = [\textbf{a}, u_{r+2},...,u_n].
\end{equation}
In conclusion, we can formulate the SIT-$\ell_1$ minimization for problem ($\ref{DualForm2}$) as below:
\begin{equation}
\begin{split}
\label{FinalProblemForRandomize}
\min_{e} \|e\|_1, s.t. \ {\textbf{a}}^T e =t, \ [u_{r+2},...,u_{n}]^T e =0.
\end{split}
\end{equation}

We have verified that SIT-$\ell_1$ minimization can be applied to obtain a sparsest solution to an under-determined system in the last section. The next step is to present how to apply a randomized algorithm to solve the SIT-$\ell_1$ minimization problem.
The basic idea of the randomized algorithm is as follows: First we generate many random vectors $\textbf{a}$. For each generated random vector $\textbf{a}$, we can adopt a fast and accurate first-order method \cite{ becker2011nesta}, the linear programming method \cite{boyd2004convex}, or many other algorithms to obtain a solution to problem ($\ref{FinalProblemForRandomize}$).
Ignoring differences in magnitude, the sparsest one of all these solutions can be taken approximately to be the final sparsest error vector. If the number of random vectors $\textbf{a}$ is large enough,
then the event that the randomized algorithm will eventually return a truly sparsest error vector happens with a high probability.

Algorithm 1 presents the framework of the proposed randomized algorithm for the sparsest error detection. This program was implemented via Matlab.
\begin{algorithm}[h]
\caption{\label{alg1} A Randomized Algorithm for Sparsest Detection}
\begin{algorithmic}[1]
\REQUIRE  $y\in R^m$, $A\in R^{m\times n}$, $\epsilon$ and $SNbr$
\ENSURE
\STATE Compute $u_{r+1}, U_{r}, U_{r+1}^c$: \\
\quad $U_r = U_1(:,1:r)$ where $[U_1,S_1,V_1] =svd(A)$;\\
\quad $u_{r+1} =\frac{\tilde{u}_{r+1}}{\|\tilde{u}_{r+1}\|_2}$ where $ \tilde{u}_{r+1} = y - U_r(U_r^T U_r)^{-1} U_r^T y$;\\
\quad $U_{r+1}^c = U_2(:,r+2:m) $, $[U_2, S_2, V_2 ]= svd([A,y])$. \\
\STATE  Compute $y^*$:\\
\quad $y^* = [t,0,...,0]^T\in R^{m}$ where $t = u_{r+1}^T y$.\\
\STATE Randomly Generate $\textbf{a}$: \\
 \quad generate $\tilde{f}\in R^{r+1}$ by i.i.d Gaussian distribution;\\
\quad $\textbf{a} =  U_{r+1}f$ where $f= \frac{\tilde{f}}{\|\tilde{f}\|_2}$.\\
\STATE Adopt Basis Pursuit Algorithm:\\
\quad $\textbf{for} \ {i = 1\ :\ SNbr}$ \\
\quad\quad \quad  generate $\textbf{a}$ by step 3;\\
\quad\quad \quad set $F = [\textbf{a}, U_{r+1}^c]^T$;\\
\quad\quad \quad $\hat{e}_i = \arg\min_{e} \|e\|_1,  s.t.\ y^* = Fe $;\\
\quad\quad \quad $\hat{e}_i = sign(\hat{e}_i)\circ \left[ |\hat{e}_i| -\epsilon \vec1\right]_{+}$;\\
\quad\quad\quad $i = i+1$.\\
\quad\textbf{end for}\\
\STATE Find the sparest solution $\hat{e}$:\\ 
\quad $\hat{e} = \arg \min_{e\in \{\hat{e}_1,...,\hat{e}_{SNbr}\}} \|e\|_0$.\\
\RETURN $\hat{e}$
\end{algorithmic}
\end{algorithm}
In algorithm 1, "SNbr" represents the number of the randomly generated values for $\textbf{a}$, the operation "svd(A)" implements the singular value decomposition on $A$, $U(:,n:m)$ denotes a sub-matrix formed by selecting columns from $n$ to $m$ of $U$, $\left[ x \right]_{+}$ denotes a new vector achieved by setting the negative entries of $x$ to zeros whereas the positive entries remain the same,
and the notation " $\circ$ " denotes the Hadamard product, i.e., the matrix obtained by entry-by-entry multiplication.

\begin{figure}[t]
\begin{center}
   \includegraphics[width=12cm,height=14cm]{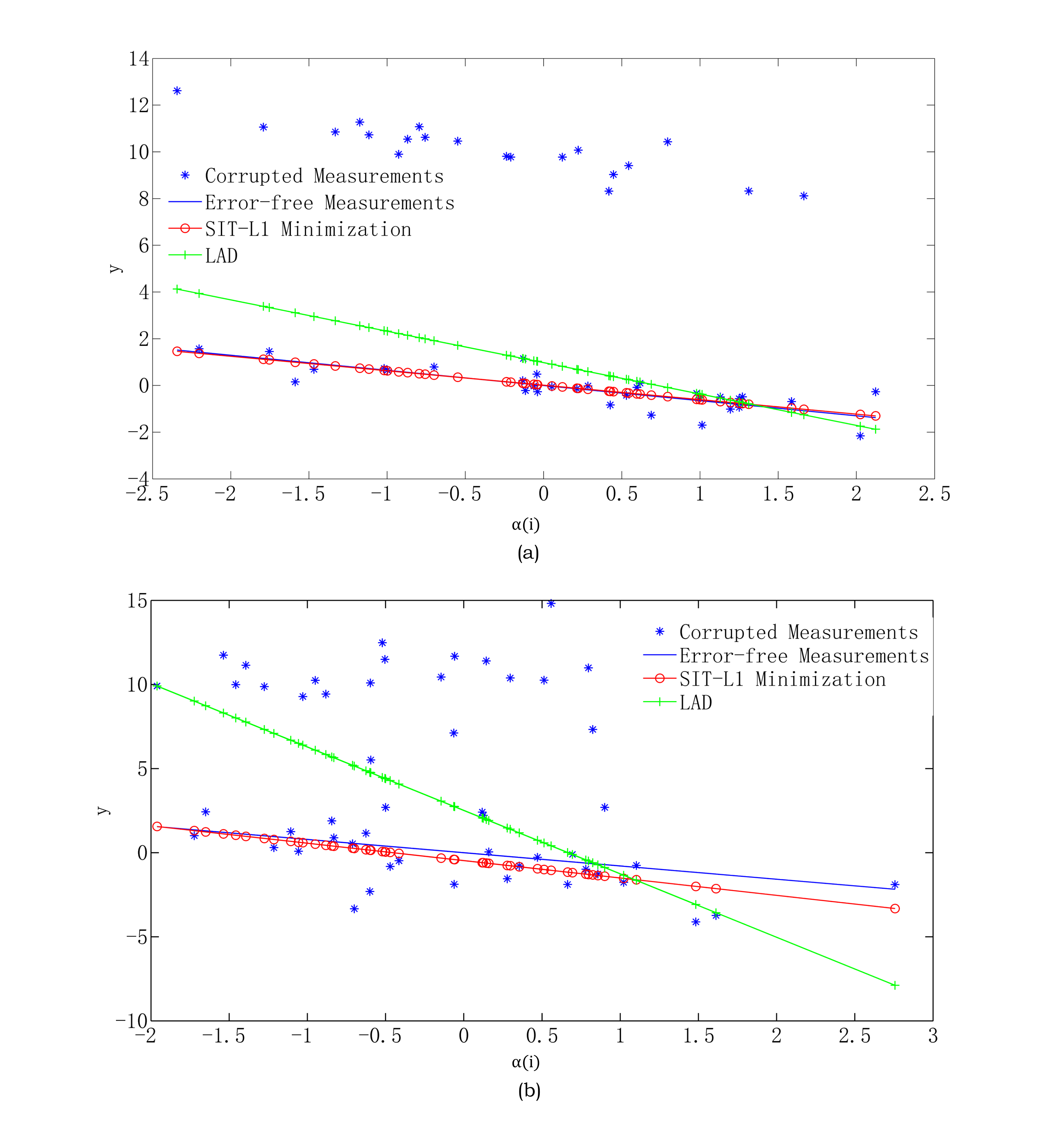}
\end{center}
\caption{Comparing Performance with a Uniform Outlier Distribution: Estimations of $y$ obtained from different algorithms are compared against the exact value. $\alpha (i)$ denotes the x-coordinate of the $i$-th blue point, whose $y$-coordinate represents the actual measurement. The blue line represents the error-free $y$ value, the red line represents the estimated value according to SIT-$\ell_1$ minimization, and the green line the estimated value achieved by LAD. The noise level is relatively small in figure (a) and relatively high in figure (b). For both cases, the estimations provided by the proposed method are more precise than those provided by LAD.}
\end{figure}

Due to the limited computational accuracy of Matlab, most entries of a computed sparse vector may not be exactly zeros.
To mitigate this problem, we set a positive threshold $\epsilon$ in algorithm 1, which is used to set entries with very small values (smaller than the $\epsilon$) to zeros.

\begin{figure}[t]
\begin{center}
   \includegraphics[width=12cm,height=7cm]{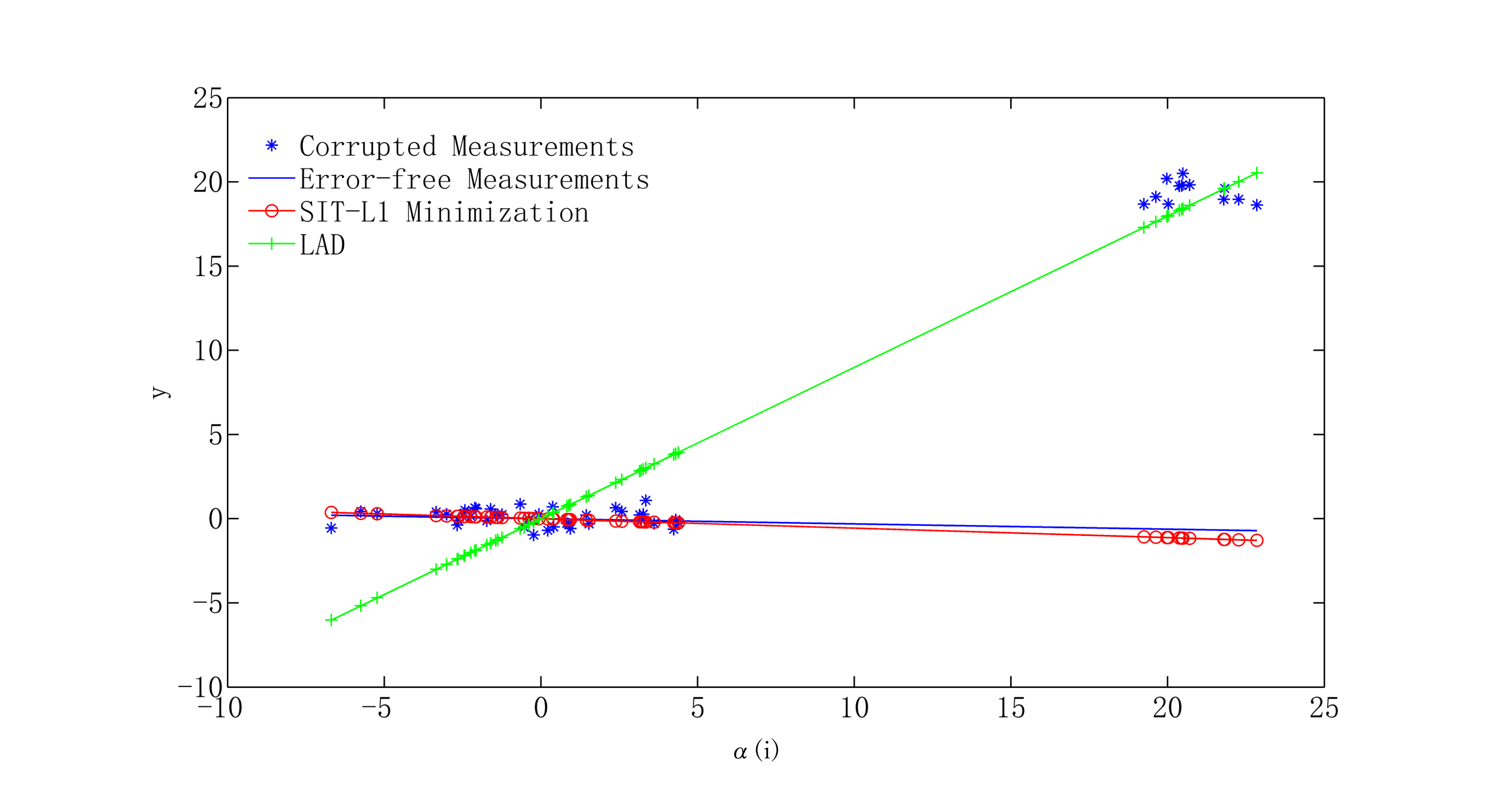}
\end{center}
\caption{Comparing Performance with a Tailored Outlier Distribution: Estimations of $y$ obtained by different algorithms are compared against the exact value. $\alpha(i)$ indicates the x-coordinate of an $i$-th blue point, whose y-coordinate represents the actual measurement. The blue line represents the error-free y value, the red line represents the estimated value according to SIT-$\ell_1$ minimization and the green line the estimated value achieved by LAD. This result suggests that the proposed method has a much higher resistance to arbitrarily distributed errors than the LAD method.}
\end{figure}

\section{Numerical Study}
In this section, we present a series of experiments to demonstrate the effectiveness of the SIT-$\ell_1$ minimization in sparsest error detection.
In the first subsection, we aim to provide readers a quick impression of the performance of the proposed method to robust linear regression.
In the second subsection, we aim to highlight the statistical performance of the proposed method for sparsest error detection involving high dimensional data sets.
In these two subsections, we use $\mathcal{N}(a,b)$ to denote a Gaussian distribution with mean $a$ and
standard deviation $b$.

\subsection{Robust Linear Regression}
Linear regression is an important methodology in statistics. It aims to find the hidden linear relationship between the outputs and the inputs. Traditionally, people apply the least mean squares to achieve an estimation for the hidden signal. However, the least mean squares method usually lacks robustness, especially in cases where there are many outliers among the measurements.  Recently, researchers are increasingly interested in robust methods, which are expected to have high resistance to outliers.

The proposed SIT-$\ell_1$ minimization is a strong candidate for robust regression. To demonstrate this, we conduct comparisons between the proposed SIT-$\ell_1$ minimization and a popular robust regression method $-$ the least absolute deviation (LAD).
In fact, the LAD method is a direct $\ell_1$ minimization method.

The setup for the robust linear regression experiments is as follows:
The entries of the hidden parameter vector $x\in R^2$ are generated by a Gaussian distribution $\mathcal{N}(0,1)$.
The measurement matrix $A$ is set as:
\begin{equation}
\nonumber
A = [\alpha, \vec{1}],
\end{equation}
where $\vec{1}\in R^{52}$ is a column vector whose entries are all ones.
In Fig.2, $\alpha\in R^{52}$ is generated as $\alpha(i) = i$ for each $i$ in $[0,56]$.
In Fig.3, $\alpha$ is generated as
\begin{equation}
\nonumber
\alpha(i)= \begin{cases}
i, & i\le 32\\
i+20,& i>32.
\end{cases}
\end{equation}
The number of the non-zero entries of the error vector, $e\in R^{52}$, is set to be $20$. Moreover, if the $i$-th entry of $e$, $e(i)$, is non-zero then $e(i)$ is set to be $20$.
To make the simulation study more realistic, we include additional small additive Gaussian noises to the measured data $y$. Therefore, the final $y$ is generated by the equation $y = Ax + e +z$ where $z$ denotes a Gaussian noise vector.
In Fig.2 $(a)$ and in Fig.3, entries of $z$ are generated by the Gaussian distribution $\mathcal{N}(0,0.5)$ and the threshold $\epsilon$ is set to be 0.3. Whereas in Fig.2 (b), entries of $z$ are generated by $\mathcal{N}(0,2)$ and $\epsilon = 1.8$. In Fig.2 and Fig.3, SNbr, the sample number, is set to 80.
The task of the corresponding linear regression problem is to estimate the hidden parameters (i.e., the entries in $x$) as accurate as possible when given $y$ and $A$. Therefore we can apply algorithm 1, which is presented in the last section for SIT-$\ell_1$ minimization, to get an estimation of the real hidden parameters.
Since Gaussian noises are included in the measurements, we replace the basis pursuit algorithm with the basis pursuit denoising algorithm \cite{gill2011crowd} in algorithm 1.

Fig.2 shows the SIT-$\ell_1$ minimization is more robust than the LAD method under different levels of Gaussian noises. Fig.3 presents a set of data based on a tailored error distribution to highlight the weakness of direct $\ell_1$ minimization. The results indicate that it totally loses the ability to distinguish outliers for the data set presented, whereas the performance of the proposed method is resistant to this change in error distribution.

\subsection{Statistical Performance of the SIT-$\ell_1$ Minimization}
It has been shown that direct $\ell_1$ minimization is efficient in detecting the sparsest errors when the measurement matrix is generated at random with \textit{i.i.d. entries} \cite{candes2005decoding, candes2008highly, xu2013sparse}. However, there are situations where direct $\ell_1$ minimization fails. SIT-$\ell_1$ minimization offers a feasible alternative for these cases. To demonstrate this, we generate a measurement matrix with entries that are random but not independently identically distributed.  For this measurement matrix, the comparison shows that direct $\ell_1$ minimization fails to detect \textit{arbitrarily-distributed} sparse errors whereas the proposed method remains effective.

In order to demonstrate this, we only need to generate a random matrix $A$ by two Gaussian distributions with different means. We set the row number of $A$ to $256$. Given an integer $k\in (0,256)$, let each entry in the first $256-k$ rows of $A$ be generated by $\mathcal{N}(0,1)$, whereas entries in the remaining rows of $A$ be generated by $\mathcal{N}(5,1)$. To test whether both SIT-$\ell_1$ minimization and  direct $\ell_1$ minimization can detect sparse errors with arbitrary distributions, we set up a series of error distributions as follows: Let $s$ denote the number of the occurring errors, that is, $\|e\|_0 =s$, given a real number $t\in [0,1]$, let $t*s$ be number of errors which appear among the first $256-k$ entries of $e$. These errors are distributed uniformly.
Similarly, let the rest of the errors appear uniformly among the last $k$ entries of $e$. We can obtain different error distributions by varying t in $[0, 1]$. Moreover, we set $e(i) = 10$ if $e(i)\neq 0$. and we generate each entry of $x$ by $\mathcal{N}(0,1)$, then the final $y$ can be realized by $y=Ax+e$. In the following experiments, unless otherwise stated, we set $\epsilon =0.2$.

\begin{figure}[t]
\begin{center}
   \includegraphics[width=12cm,height=7cm]{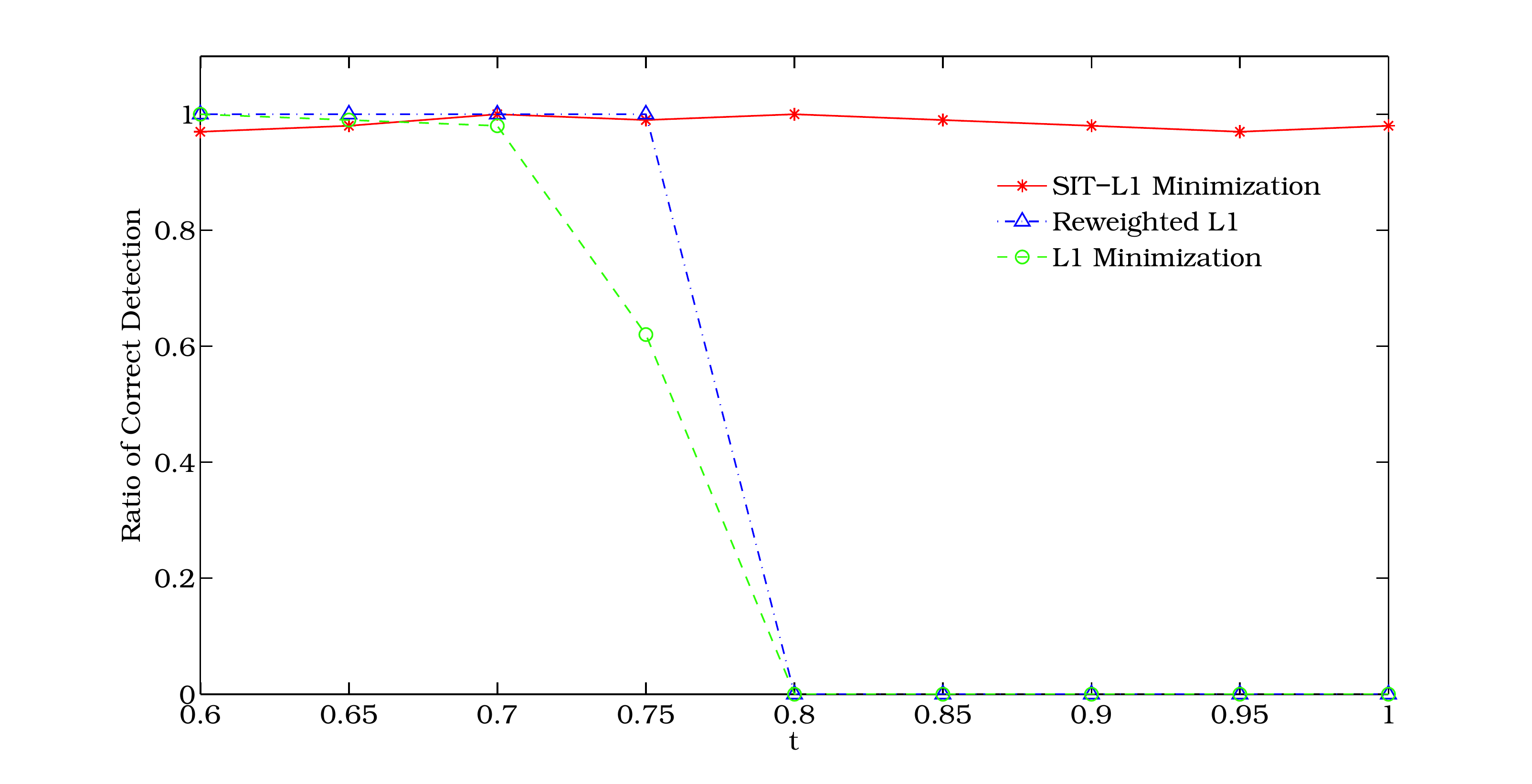}
\end{center}
\caption{Comparison under Different Error Distributions: Different error distributions were obtained by varying t from $0.6$ to $1$. The graph shows the ratio of correct detections for a given method, which is defined as the proportion of the cases the method correctly predicts whether there is an error or not in the measurement.
}
\end{figure}

\begin{figure}[t]
\begin{center}
   \includegraphics[width=12cm,height=7cm]{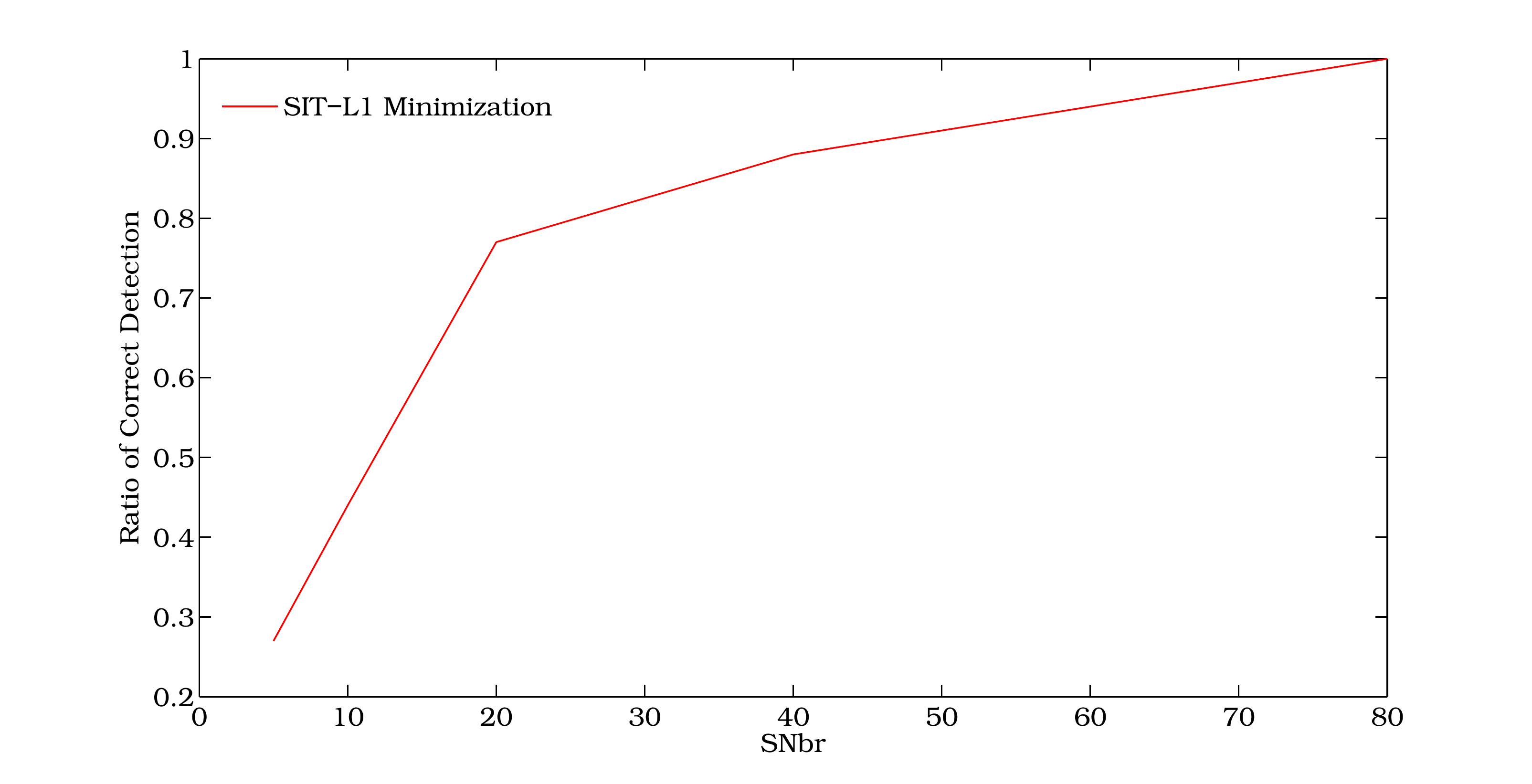}
\end{center}
\caption{Detection Accuracy versus Sample Number: The graph shows the ratio of correct detections for a given method (as defined in Fig.4) against the number of samples. A turning point appears around $SNbr =20$. Generally speaking, the higher the SNbr, the better is the performance of the proposed algorithm.}
\end{figure}

\begin{figure}
\begin{center}
   \includegraphics[width=12cm,height=7cm]{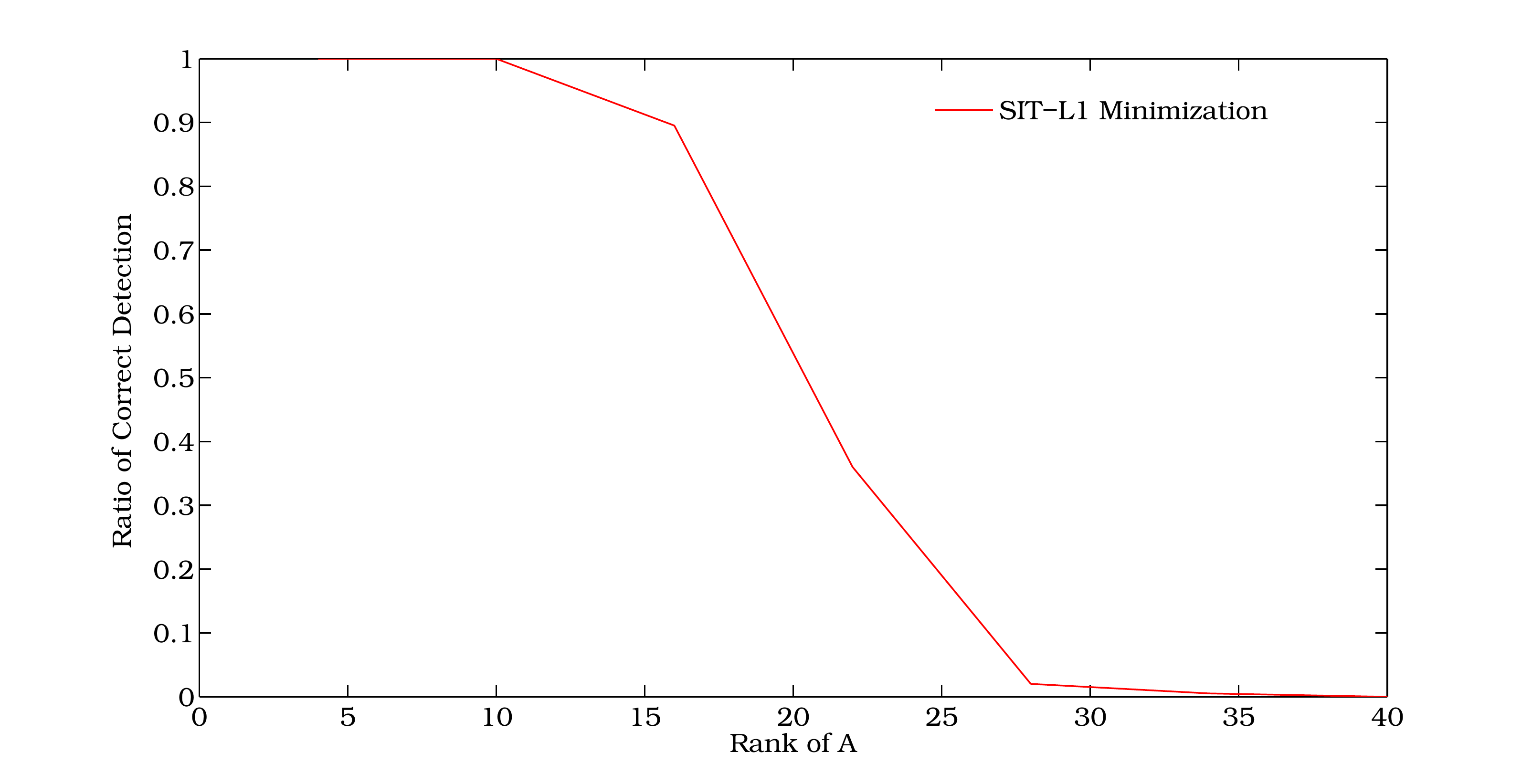}
\end{center}
\caption{Detection Accuracy versus  Rank of $A$: The graph shows the ratio of correct detections for a given method (as defined in Fig.4) against the rank of $A$. As the rank increases, the ratio of correct detection decreases.}
\end{figure}

\begin{figure}[t]
\begin{center}
   \includegraphics[width=12cm,height=7cm]{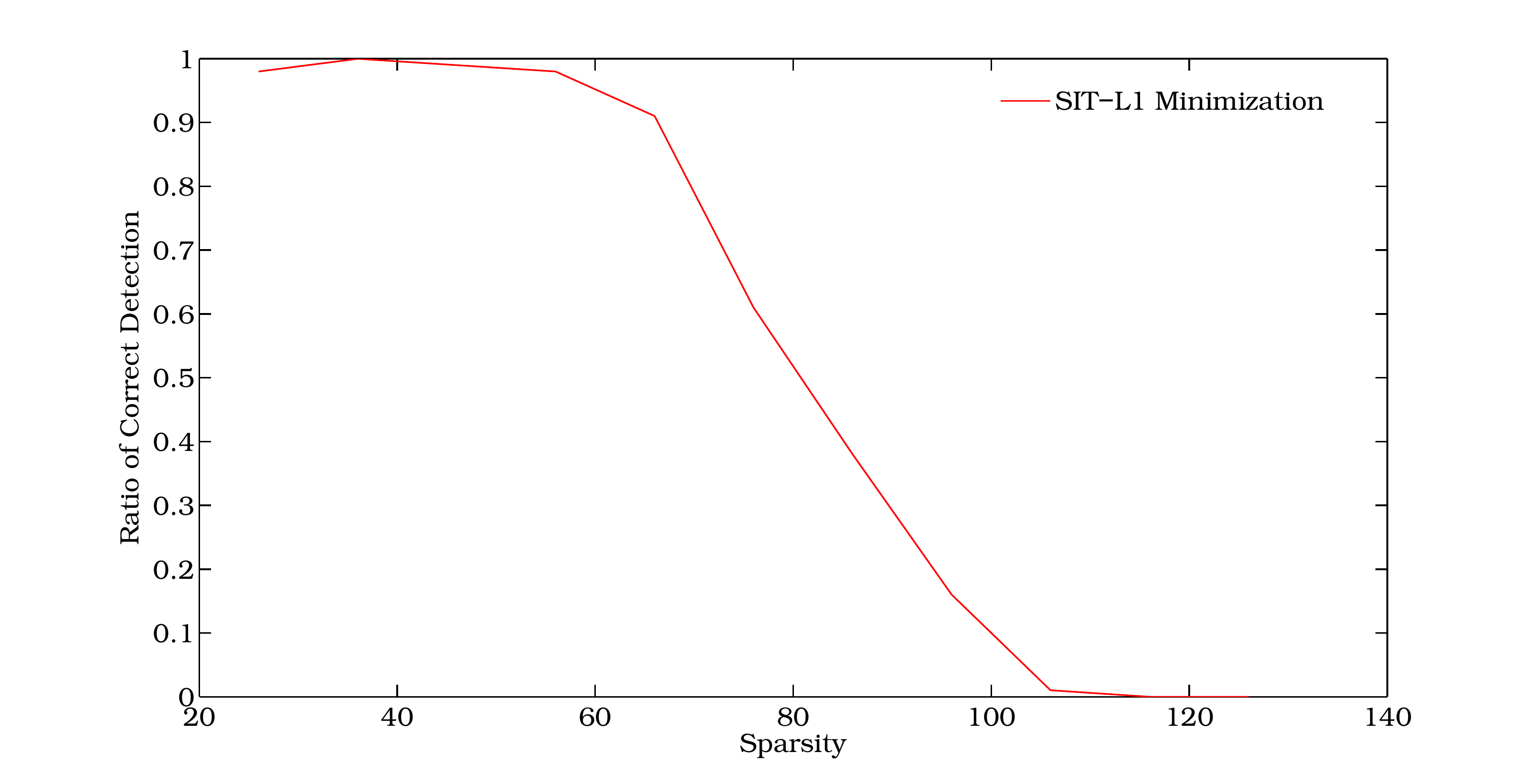}
\end{center}
\caption{Detection Accuracy Versus Number of Errors: as the error number increases, it becomes harder for the proposed method to detect all the errors correctly.}
\end{figure}

In Fig. 4, we set $s = 36$, $r =16$, ( $r$ is the rank of $A$,) $k=56$ and $SNbr =100$. The graph compares the proposed method with two popular methods, direct $\ell_1$ minimization and reweighted $\ell_1$ minimization \cite{candes2008enhancing}, under different error distributions parameterized by $t$. For each $t$, we test 200 trials and get a ratio of correct detection for each method. From Fig.4, we can see, when $t>0.75$ both direct $\ell_1$ minimization and reweighted $\ell_1$ minimization frequently fail to detect the sparse errors whereas the performance of the proposed method remains unaffected by the change of error distribution.


In below, we aim to demonstrate how the detection accuracy of the proposed method varies with $SNbr$, $r$ and $s$ respectively. For simplicity, we set $k=s$ and $t=1$. Under such a parameter setting, we observe that both the direct $\ell_1$ minimization and the reweighted $\ell_1$ minimization fail to detect all the sparse errors exactly. Hence, the following figures only present the performance of the proposed method. We test 200 trials for each value setting of $(SNbr, r, s)$.

Generally speaking, the larger the $SNbr$, the higher probability for the proposed algorithm to detect all the sparse errors. But the time to search for the solution is shorter when the value of SNbr is smaller. For Fig.5 we set $r=16$ and $s = 36$. The graph shows that if SNbr is larger than 20, then the proposed method can achieve a good performance in error detection.

In Fig.6 we set $s = 36$ and $SNbr = 40$. The graph shows that the detection accuracy of the proposed method decreases as the rank of $A$ increases. But if $SNbr$ increases with the rank of $A$, the detection accuracy may improve.

In Fig.7, we set $r =16$ and $SNbr =120$. The graph implies a very high error-detection capability of the proposed methodology since it stops to detect all the sparse errors exactly only when the corrupted measurements get close to half of the total measurements.

\section{Conclusion}
This paper presents a new methodology$-$SIT-$\ell_1$ minimization for sparsest errors detection of an over-determined linear system. The basic idea is to use a Sparsity Invariant Transformation (SIT) to transform the original linear system into a representation in which the $\ell_0$-minimal error vector coincides with the $\ell_1$-minimal error vector.
A contribution of this paper lies in showing the existence of such an SIT for a general over-determined linear system.
Moreover, this methodology can be applied to sparsest recovery of an under-determined linear system. Compared to the methodology proposed in compressed sensing, an advantage of the proposed methodology is the removal of structure constraints on the measurement matrices.
So far, there is no known efficient algorithm to construct a feasible SIT for a linear system. Instead, we provide a randomized algorithm based on the Monte Carlo simulation to search for one. The numerical results demonstrated the performance improvement of the proposed method in comparison with direct $\ell_1 $ minimization and reweighted $\ell_1$ minimization.

%


%

%
%

%
%

\ifCLASSOPTIONcaptionsoff
  \newpage
\fi



%
%
%
\bibliographystyle{IEEEtran}
\bibliography{IEEEabrv,IEEEexample}

%

%
%
%




\end{document}